# Magnetoelastic Transport-Path Reconstruction and Giant Magnetotransport Responses in a Two-Dimensional Antiferromagnet

Liu Yang,[1,2] Ming Li,[3,*] Shui-Sen Zhang,[4] Hang Zhou,[1,2] Yi-Dong Liu,[1,2] Xiao-Yan Guo,[1] Wen-Jian Lu,[1] Yu-Ping Sun,[4,1,5] Evgeny Y. Tsymbal,[6] Kaiyou Wang,[7,8,†] and Ding-Fu Shao[1,‡]

[1] *Key Laboratory of Materials Physics, Institute of Solid State Physics, HFIPS, Chinese Academy of Sciences, Hefei 230031, China*
[2] *University of Science and Technology of China, Hefei 230026, China*
[3] *School of Physics, University of Electronic Science and Technology of China, Chengdu 611731, China*
[4] *Anhui Key Laboratory of Low-Energy Quantum Materials and Devices, High Magnetic Field Laboratory, HFIPS, Chinese Academy of Sciences, Hefei 230031, China*
[5] *Collaborative Innovation Center of Microstructures, Nanjing University, Nanjing 210093, China*
[6] *Department of Physics and Astronomy & Nebraska Center for Materials and Nanoscience, University of Nebraska, Lincoln, Nebraska 68588-0299, USA*
[7] *State Key Laboratory of Semiconductor Physics and Chip Technologies, Institute of Semiconductors, Chinese Academy of Sciences, Beijing 100083, China*
[8] *Center of Materials Science and Optoelectronics Engineering, University of Chinese Academy of Sciences, Beijing 100049, China*

[*] ming.li7@uestc.edu.cn; [†] kywang@semi.ac.cn; [‡] dfshao@issp.ac.cn

Nonvolatile magnetotransport responses in a single magnetic material have generally not been expected to exhibit a large ON/OFF ratio, because they are usually tied to spin-orbit coupling and therefore remain relatively weak. Here we show, contrary to this expectation, that giant nonvolatile magnetotransport can arise in a single magnetic material through magnetoelastic reconstruction of nonrelativistic real-space transport paths. Using the two-dimensional antiferromagnet FePS$_3$ as a representative system, first-principles quantum transport calculations reveal that charge transport is strongly tied to its quasi-one-dimensional zigzag sublattice chains and, under suitable doping, can even become confined to them. Moreover, strain lifts the degeneracy among symmetry-related zigzag variants and thus reorients these transport paths through magnetoelastic coupling. As a result, both the longitudinal and transverse conductivities change dramatically, yielding a giant magnetoelastic magnetoresistance of up to $10^4$% and an energy-independent Hall ratio that far exceeds the spontaneous Hall ratios found in conventional magnets. These results establish a route to exploiting symmetry-related magnetic variants and their associated transport paths for reconfigurable, high-performance spintronic devices with large nonvolatile readout contrast.

In magnetic materials, transport is shaped by the interplay among electronic structure, magnetic order, symmetry, and scattering, giving rise to a broad range of spin-dependent responses. Depending on how magnetic order couples to charge motion, these responses can appear in the longitudinal channel as magnetoresistance (MR) [1-9] and in the transverse channel as Hall effects [10-12]. Beyond their rich physical content, such effects are especially important because they provide electrical access to magnetic states and therefore form a basic foundation for spintronic functionalities [13].

For realistic information applications, however, the central requirement is not merely the existence of a transport signal, but the ability to electrically distinguish stable, ideally nonvolatile, magnetic states. The most successful strategies so far have largely relied on multilayer magnetic heterostructures, exemplified by giant MR (GMR) [2,3,14] and tunneling MR (TMR) [7-9]. These approaches have enabled practical memory and readout technologies, but they also depend on complex layer stacks, interfacial engineering, and demanding fabrication control [15]. By contrast, intrinsic transport responses in a single magnetic material, such as anomalous Hall effect (AHE) [12] and anisotropic magnetoresistance (AMR) [1,16,17], are in principle more attractive for device simplification because they do not require multilayer junction architectures. Yet in most known systems, their switching amplitudes remain too limited to provide robust high-contrast nonvolatile readout comparable to heterostructure-based devices.

This contrast raises a fundamental question: can one realize a large nonvolatile magnetotransport response in a single magnetic material? Addressing this challenge requires going beyond the conventional picture in which transport is modulated only by the reorientation of magnetic moments. In a single magnetic material, such responses are primarily tied to magnetic anisotropy induced by spin-orbit coupling (SOC), which usually limits the attainable signal amplitude. As a result, although devices based on a single magnetic material are structurally simpler, their intrinsic transport responses have rarely provided a competitive route toward high-contrast nonvolatile readout.

A different opportunity may emerge in antiferromagnets – ferroic materials which have two or more magnetic sublattices. Although antiferromagnets were long regarded as essentially spin-independent in transport, they possess several key advantages for spintronics, including zero stray fields, insensitivity to external magnetic-field perturbations, and ultrafast dynamics [18]. Leveraging these features, recent studies based on macroscopic symmetry and momentum-space electronic structure have uncovered a wide range of functionalities of antiferromagnets useful for spintronics [19-38]. In addition to these established directions, antiferromagnets may host another, less explored opportunity rooted in their real-space sublattice structure. Recent works suggest that transport can be shaped not only by the orientation of magnetic moments, but also by the geometry and connectivity of sublattice-resolved transport paths [39-44]. For particular antiferromagnetic (AFM) stackings, spin-dependent transport may even become confined to low-dimensional channels associated with specific sublattices [39, 40]. This points to a mechanism qualitatively distinct from conventional SOC-driven AMR: instead of weakly tuning the conductivity tensor through moment rotation, one may generate a much stronger response by reconstructing the spin-dependent transport paths themselves. If such highly anisotropic sublattice-associated transport paths can be externally switched between distinct nonvolatile configurations, a large transport contrast may then become possible even in a single magnetic material.

Here we theoretically demonstrate this mechanism in a two-dimensional (2D) antiferromagnet with zigzag magnetic order. Using monolayer $FePS_3$ as a representative example, we show that doped carriers are confined to sublattice-selective quasi-one-dimensional transport paths, and that these paths can be reoriented by controlling the zigzag magnetic variant through magnetoelastic coupling. As a result, the Néel spin currents are redirected between distinct nonvolatile transport paths, leading to pronounced switching of both longitudinal and transverse transport responses. Specifically, the longitudinal MR can reach an extraordinary magnitude of ~$10^4$% under electron doping. The transverse response, meanwhile, exhibits a strain-switchable Hall-like behavior, with a Hall ratio that remains energy independent and far exceeds the spontaneous Hall ratios found in conventional magnets. These results establish magnetoelastic transport-path reconstruction as a route to giant magnetotransport responses in a single magnetic material, and suggest a general strategy for nonvolatile spintronic functionalities beyond conventional SOC-driven anisotropic transport.

We consider a collinear antiferromagnet consisting of two oppositely magnetized ferromagnetic (FM) sublattices (indexed by A and B). In such an antiferromagnet, real-space local transport can be understood in terms of interatomic hopping between and within magnetic sublattices [39]. The strength of such local transport is controlled mainly by two factors: the geometric separation between neighboring atoms and the spin polarization of the local density of states. Shorter interatomic distances generally enhance hopping, whereas larger separation suppresses it. At the same time, in the absence of spin-orbit coupling, spin remains a good quantum number, so spin-flip hopping is forbidden. Qualitatively reminiscent of Julliére's picture [4], the effective conductance between two sites is determined by the overlap of the spin-resolved available states at both ends; therefore, stronger sublattice spin polarization enhances both the spin selectivity and the magnitude of the intra-sublattice conductance $g_{AA}$ and $g_{BB}$, but suppresses the inter-sublattice conductance $g_{AB}$. As a result, the intra-sublattice conductances $g_{AA}$ and $g_{BB}$ are spin polarized with opposite spin characters on the two sublattices, whereas the inter-sublattice conductance $g_{AB}$ is spin neutral. In AFM structures with FM-ordered sublattices forming highly anisotropic low-dimensional geometries, such as chain-like or layer-like motifs, this naturally leads to direction-selective transport. When the electric field is applied parallel to these low-dimensional sublattices, transport is dominated by the intra-sublattice currents $J_{AA}$ and $J_{BB}$, resulting in Néel spin currents embedded in the global charge current [39]. For the perpendicular direction, transport is governed mainly by the spin-independent inter-sublattice current $J_{AB}$. Since these local current components differ markedly in both magnitude and spin character, nonvolatile switching of the orientations of such low-dimensional sublattice motifs would thus produce large magnetotransport responses in a single antiferromagnet [45].

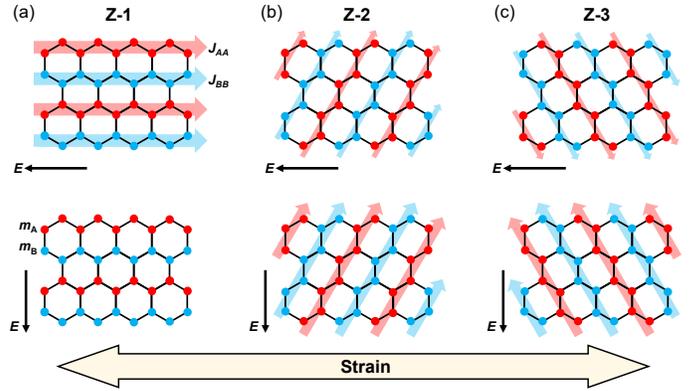

**Fig. 1:** (**a–c**) Schematic illustration of magnetoelastic reconstruction of transport paths in the three symmetry-related zigzag AFM variants Z-1 (**a**), Z-2 (**b**), and Z-3 (**c**). The zigzag chains are oriented along $0°$, $+60°$, and $-60°$, respectively, relative to the horizontal direction. Red and blue circles denote the oppositely magnetized sublattices $m_A$ and $m_B$. Red and blue arrows represent the spin-polarized intra-sublattice currents $J_{AA}$ and $J_{BB}$, respectively, where the arrow widths indicate the current magnitude, and the black arrows denote the applied electric field $E$. Strain lifts the degeneracy among the three variants and switches the corresponding real-space transport paths. For simplicity, the spin-neutral inter-sublattice current $J_{AB}$ is omitted.

This physical picture has been already exemplified in several three-dimensional AFM stackings that support Néel spin currents, such as A-, C-, and X-type orders [39, 40]. However, in these configurations, the transport-path orientation is structurally fixed and difficult to switch in a controllable and nonvolatile manner. While they provide important prototypes of the underlying transport mechanism, they are less suitable as platforms for switchable path reconstruction.

A more promising route is offered by antiferromagnets in which the relevant sublattice couplings are confined within the plane. A prototypical example is the zigzag AFM order on a honeycomb lattice, where zigzag FM chains run along one in-plane direction and are AFM-coupled to neighboring chains (Fig. 1(a)). Due to the threefold rotational symmetry of the honeycomb lattice, two additional degenerate zigzag variants exist, with the chain direction rotated by $+60°$ and $-60°$ (Fig. 1(b, c)). These symmetry-related magnetic variants define distinct real-space transport paths as long as the magnetic atoms host finite local spin polarizations. If an external perturbation can selectively lift their degeneracy and stabilize one variant over the others, then nonvolatile switching among different transport paths becomes possible. Strain provides such a route: by modifying the intra-chain and inter-chain distances, it changes the relative strengths of the FM and AFM exchanges and can therefore select a preferred zigzag orientation through magnetoelastic coupling. The resulting switching between different zigzag variants reconstructs the transport paths in real space and gives rise to pronounced longitudinal and transverse magnetotransport responses (Fig. 1).

To validate the above mechanism, we perform first-principles electronic-structure and quantum-transport calculations [46] on monolayer $FePS_3$, a 2D AFM semiconductor with a zigzag AFM ground state and a Néel temperature of about 120 K [47, 48]. Experimentally, $FePS_3$ can be exfoliated down to the monolayer limit [49-51]. Figure 2(a) shows the crystal and magnetic structure of monolayer $FePS_3$ for the zigzag variant whose zigzag chains run along the $x$ direction (denoted as Z-1). The Fe atoms form a slightly distorted honeycomb network, leading to an orthorhombic cell. Figure 2(b) shows the electronic band structure of Z-1 together with the corresponding density of states. Due to two magnetic sublattices being connected by the symmetry combining spatial inversion and time reversal operations, all bands are spin degenerate. A pronounced anisotropy is nevertheless immediately evident from the dispersions along $\Gamma - X$ and $\Gamma - Y$, indicating different conductivities parallel and perpendicular to the zigzag chains. This contrast is particularly strong in the conduction bands, where the dispersion along $\Gamma - X$ is much stronger than that along $\Gamma - Y$. The origin of this anisotropy is further clarified by the local density of states on the $Fe_A$ sublattice (Fig. 2(c)). In the valence bands where 3d orbitals from magnetic Fe atoms and 2p orbitals from nonmagnetic S atoms hybridize, both spin channels contribute, leading to a relatively weak local spin polarization (Figs. 2(c) and S2). In contrast, the conduction bands are dominated by the Fe atoms and host a single spin channel, implying strong local spin polarization (Figs. 2(c) and S2). According to the transport picture discussed above, this favors strong intra-sublattice conductances $g_{AA}$ and $g_{BB}$ while strongly suppressing the inter-sublattice conductance $g_{AB}$, thereby promoting transport along the zigzag chains and weakening it in the transverse direction.

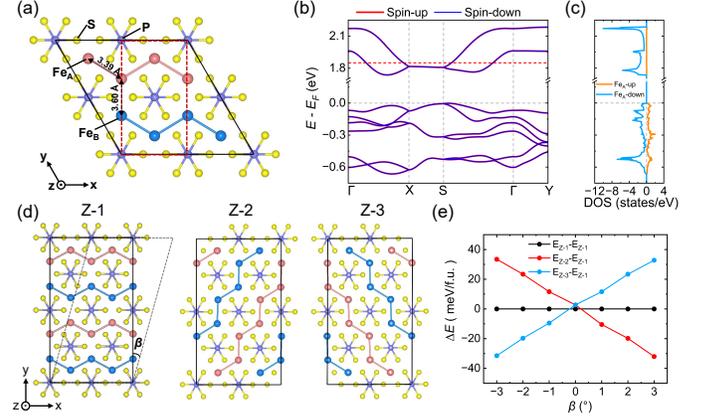

**Fig. 2:** **(a)** Crystal and magnetic structure of monolayer $FePS_3$ in the Z-1 zigzag AFM variant. Fe atoms form a slightly distorted honeycomb lattice, giving rise to an orthorhombic unit cell. **(b,c)** Electronic band structure **(b)** and $Fe_A$-resolved projected density of states **(c)** of Z-1. **(d)** The orthorhombic supercells for the three symmetry-related zigzag variants Z-1, Z-2, and Z-3. The shear strain is parameterized by the shear angle $\beta$. **(e)** Relative energies of the three zigzag variants as functions of $\beta$, demonstrating pronounced magnetoelastic coupling and strain-selective stabilization of different zigzag variants.

A further key feature of monolayer $FePS_3$ is that the Fe-Fe distances within and between zigzag chains are different: the intra-sublattice Fe-Fe distance is 3.39Å, whereas the inter-sublattice separation is 3.60Å. This difference suggests that the competing exchange interactions are sensitive to lattice distortion, opening a route to controlling the relative stability of different zigzag variants by strain. To verify this idea, we construct the three symmetry-related zigzag variants, with the rotated variants denoted as Z-2 and Z-3, in a common orthorhombic cell, as shown in Fig. 2(d). Instead of applying direct uniaxial strain along each zigzag direction, which is inconvenient computationally for Z-2 and Z-3 variants, we apply a shear strain along $x$, parameterized by the shear angle $\beta$. The positive (negative) $\beta$ thus stretches the zigzag chains for Z-2 (Z-3). As shown in Fig. 2(e), the three variants are nearly degenerate at $\beta = 0$, with only a tiny residual splitting caused by forcing the rotated variants into the same orthorhombic cell. Remarkably, even a small shear strain lifts this degeneracy strongly: positive $\beta$ stabilizes Z-2, whereas negative $\beta$ stabilizes Z-3. The large energy splitting suggests that such strain is sufficient to overcome the barrier between competing variants, making nonvolatile switching between zigzag variants feasible. These results demonstrate pronounced magnetoelastic coupling and indicate that strain can be used to switch among the three zigzag variants, consistent with a recent experiment on a similar zigzag antiferromagnet FePSe$_3$ [52].

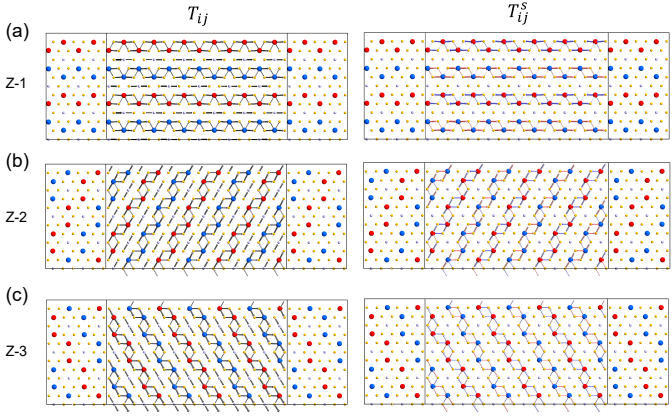

**Fig. 3:** (**a–c**) Real-space bond charge transmission $T_{ij}$ (left) and bond spin transmission $T_{ij}^s$ (right) for electron-doped monolayer FePS$_3$ in the three zigzag variants Z-1 (**a**), Z-2 (**b**), and Z-3 (**c**), evaluated at the energies marked by the red dashed lines in Fig. 2(b). The dominant charge transmission is strongly confined to the zigzag chains, while the bond spin transmission shows highly spin-polarized chain transport with opposite spin flow on neighboring magnetic sublattices, evidencing the pronounced Néel spin currents. Arrow width denotes the magnitude of $T_{ij}$ or $T_{ij}^s$; in the right panels, red and blue indicate opposite spin-transport directions. Spin-up and spin-down magnetic moments are shown in red and blue, respectively.

The magnetoelastic reconstruction has a strong effect on transport paths. To demonstrate this, we consider transport along the $x$ direction and calculate the bond-resolved charge and spin transmissions [53] defined as $T_{ij} = T_{ij}^\uparrow + T_{ij}^\downarrow$ and $T_{ij}^s = T_{ij}^\uparrow - T_{ij}^\downarrow$, respectively, where $i$ and $j$ label the two atomic sites connected by a given bond, and $\uparrow$ and $\downarrow$ denote the two spin channels. We focus on n-doped FePS$_3$ at the energies marked by the red dashed lines in Fig. 2(b), since the conduction bands of monolayer FePS$_3$ exhibit a much stronger spin polarization than the valence bands. Figure 3 shows the real-space bond charge transmission $T_{ij}$ and bond spin transmission $T_{ij}^s$ for the three zigzag variants. In all three cases, the bond charge transmission is strongly concentrated on the zigzag chains, directly visualizing that the dominant transport paths are confined along the chain direction rather than spreading isotropically over the 2D lattice (Fig. 3(a)). Moreover, when the zigzag orientation changes from Z-1 to Z-2 and Z-3, the dominant transport paths rotate accordingly in real space, consistent with the transport-path reconstruction picture discussed above.

The bond spin transmission further reveals that the chain-confined transport is highly spin polarized (Fig. 3(b)). In each variant, the dominant spin transmission is carried by the intra-chain bonds on the two magnetic sublattices, with opposite spin polarizations on neighboring zigzag chains. This indicates that, under electron doping, transport is dominated by the spin-polarized intra-sublattice currents $J_{AA}$ and $J_{BB}$, while the spin-neutral inter-sublattice current $J_{AB}$ is suppressed. Therefore, Fig. 3 not only confirms that electron transport in n-doped FePS$_3$ is confined to zigzag paths, but also shows that the reconstructed transport paths simultaneously carry strong Néel spin currents.

In real materials, disorder and defects affect electronic transport. We therefore go beyond the local ballistic picture and evaluate the diffusive conductivity tensor based on the relaxation time approximation using the first-principles-derived tight-binding Hamiltonian. The resulting longitudinal and transverse conductivities are shown in Fig. 4.

We first discuss the longitudinal response. Along the $x$ direction, the conductivity $\sigma_{xx}$ is maximal for Z-1, while it is strongly reduced for Z-2 and Z-3 (Fig. 4(a)). As a result, the magnetoelastic MR ratio between the high- and low-conductivity zigzag variants for a given current direction, defined as $MR = \frac{\sigma_{ii}^{high} - \sigma_{ii}^{low}}{\sigma_{ii}^{low}}$, reaches the order of 100% in this direction. Notably, Z-2 and Z-3 exhibit identical longitudinal conductivities, because their zigzag transport paths are related by mirror-equivalent rotations of $+60°$ and $-60°$ with respect to the applied field, giving the same longitudinal projection. This behavior is fully consistent with the transport-path picture in Fig. 3: for Z-1, the dominant spin-polarized currents $J_{AA}$ and $J_{BB}$ flow parallel to the applied field, whereas for Z-2 and Z-3 the same transport paths are rotated away from the field direction by $\pm 60°$, thereby reducing their longitudinal projection.

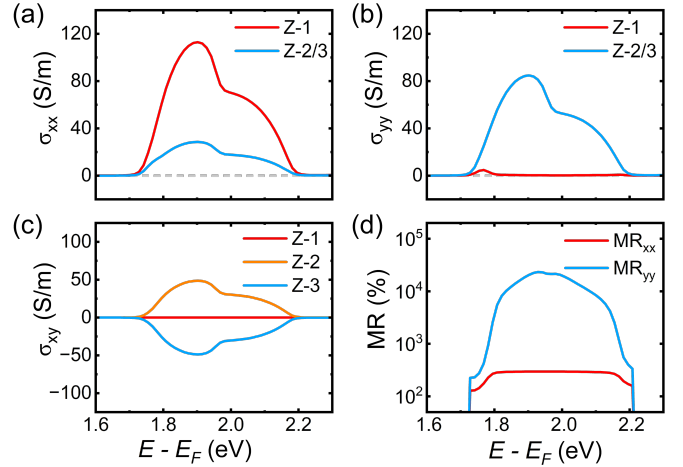

**Fig. 4:** (**a-d**) Longitudinal conductivities $\sigma_{xx}$ (**a**) and $\sigma_{yy}$ (**b**), transverse conductivity $\sigma_{xy}$ (**c**), and magnetoelastic MR ratio (**d**) as functions of energy for the electron-doped FePS$_3$. The conductivity tensor varies strongly among the three zigzag variants, consistent with the picture of the magnetoelastic reconstruction of the transport paths.

The effect is even more dramatic along the $y$ direction (Fig. 4(b)). For Z-1, transport along $y$ is governed mainly by the inter-sublattice current $J_{AB}$, with essentially no contribution from the spin-polarized intra-sublattice currents $J_{AA}$ and $J_{BB}$. Because the strong spin polarization of the Fe states strongly suppresses $J_{AB}$, $\sigma_{yy}$ of Z-1 becomes extremely small. By contrast, for Z-2 and Z-3 the dominant intra-sublattice transport paths still retain finite projections along $y$, producing much larger $\sigma_{yy}$ and a giant MR ratio reaching the order of $10^4\%$. For the same symmetry reason, Z-2 and Z-3 also have identical $\sigma_{yy}$.

The transverse response provides an equally clear signature of transport-path reorientation. As shown in Fig. 4(c), $\sigma_{xy}$ vanishes for Z-1, but becomes finite for Z-2 and Z-3 with opposite signs. Unlike a conventional anisotropy-induced transverse response, which remains fixed for a given current direction, switching between Z-2 and Z-3 here reverses the sign of $\sigma_{xy}$, producing a Hall-like response tied directly to transport-path reorientation. This sign reversal follows naturally from the real-space transport picture: under an electric field along $x$, the dominant transport paths in Z-2 and Z-3 are reoriented at $+60°$ and $-60°$, respectively, giving rise to opposite transverse currents. Although these magnetoelastic Hall conductivity itself varies with energy, the Hall ratio, defined as $|\sigma_{xy}/\sigma_{xx}|$, remains essentially constant and is determined solely by the angle between the zigzag transport path and the applied electric field. In the present case, transport-path deflection by $\pm 60°$ gives a constant Hall ratio of $\tan 60° = \sqrt{3}$, far exceeding the spontaneous Hall ratios found in conventional magnetic materials [12].

For completeness, we also examine the p-doped regime in the Supplementary Material. Because the local spin polarization near the valence bands is much weaker, all three current components, $J_{AA}$, $J_{BB}$, and $J_{AB}$, contribute to transport. Nevertheless, the bond-transmission analysis still shows visible transport-path reconstruction, and the diffusive calculation yields an magnetoelastic MR ranging from a few tens to a few hundreds of percent, far exceeding conventional AMR and comparable to typical GMR and TMR values. This indicates that, although strong sublattice spin polarization maximizes the effect, the mechanism remains robust even for more moderate spin polarization.

Since pristine FePS$_3$ is semiconducting and therefore does not exhibit the predicted response at the Fermi level, carrier doping is needed to activate the effect. In 2D materials, this should be experimentally accessible through electrostatic gating, which provides a convenient and reversible means to tune the Fermi level into the relevant bands [54].

The essential role of strain here is to lift the degeneracy among symmetry-related magnetic variants. In principle, the same function could also be served by other perturbations capable of selectively stabilizing one variant over another, such as electric or magnetic fields, temperature gradient, etc., depending on the material. It would also be interesting to explore whether spin torque [55], usually invoked for moment switching, can be adapted to switch between such transport-path variants.

We note that such a mechanism is typically absent in conventional single-sublattice ferromagnets, which generally lack the multiple magnetic sublattices needed to support reconfigurable, sublattice-resolved transport paths. More generally, the mechanism is not restricted to zigzag antiferromagnets, but should also apply to antiferromagnetic or ferrimagnetic systems with magnetic variants related by rotations about an out-of-plane axis, provided that the relevant magnetic sublattices are low-dimensional and that both their orientations and inter-sublattice antiferromagnetic couplings lie in plane. We point out, however, that antiferromagnets are expected to exhibit particularly strong magnetoelastic magnetotransport responses because of their perfectly staggered sublattice spin polarizations. In this sense, this work suggests a distinct direction for antiferromagnetic spintronics by exploiting functionalities genuinely intrinsic to antiferromagnets.

Overall, our results highlight the importance of sublattice geometry and orientation as an independent degree of freedom for controlling spin-dependent transport in antiferromagnets. Beyond the conventional focus on spin-orbit-coupling-driven moment reorientation, we show that the real-space arrangement of low-dimensional magnetic sublattices can directly govern how charge and spin currents propagate, and that switching between symmetry-related sublattice configurations can generate exceptionally strong nonvolatile magnetotransport responses in a single material. This perspective elevates sublattice-resolved transport from a microscopic description to a practical design principle for spintronic functionalities. Our work thus suggests a route toward exploiting symmetry-related sublattice orientations and their associated transport paths for reconfigurable, high-performance spintronic devices.

**Acknowledgments.** This work was supported by the National Key R&D Program of China (Grant No. 2024YFB3614100), the National Natural Science Foundation of China (Grants Nos. 12241405, 12274411, and 12504098), the Basic Research Program of the Chinese Academy of Sciences Based on Major Scientific Infrastructures (Grant No. JZHKYPT-2021-08), and the CAS Project for Young Scientists in Basic Research (Grant No. YSBR-084). The calculations were performed at Hefei Advanced Computing Center.